# Microscopic mechanism of structural and volume relaxation below glass transition temperature in a soda-lime silicate glass revealed by Raman spectroscopy and its first principle calculations

**Taisuke Suzuki[1], Yuya Hamada[1,2], Masahiro Shimizu[1,*], Shingo Urata[2], Yasuhiko Shimotsuma[1], Kiyotaka Miura[1]**

[1] Department of Material Chemistry, Graduate School of Engineering, Kyoto University, Katsura, Nishikyo-ku, Kyoto 615-8510, Japan

[2] AGC Inc., Yokohama, Kanagawa 230-0045, Japan

*corresponding author: shimizu.masahiro.3m@kyoto-u.ac.jp

**Abstract**

To elucidate the atomistic origin of volume relaxation in soda-lime silicate glass annealed below the glass transition temperature ($T_g$), the experimental and calculated Raman spectra were compared. By decomposing the calculated Raman spectra into a specific group of atoms, we found that the Raman peak at 1050 cm$^{-1}$ corresponds to bridging oxygen with a small Si-O-Si bond angle. The experimental Raman spectra indicated that, during annealing below $T_g$, a homogenization reaction $Q_2 + Q_4 \rightarrow 2Q_3$ proceeds in the early stage of structural relaxation. Then, the Si-O-Si units with relatively small angles decrease even in the later stages, which is first evidence of ring deformation causing volume relaxation of soda-lime silicate glass because decreasing small Si-O-Si angles corresponds to the reduce of acute O-O-O angle in a ring and can expand the space inside the rings, and Na can be inserted into the ring center. In conclusion the ring deformation and Na displacement is the origin of the volume relaxation of soda-lime silicate glass below $T_g$.

## I. INTRODUCTION

Structural relaxation below the glass transition temperature ($T_g$) has attracted significant attention in both science and the industry. From a scientific viewpoint, structural relaxation below $T_g$, where the viscosity reaches a value of $10^{12}$ Pa·s in silicate glass by convention, can be interpreted as a metastable glassy state; therefore, the system will seek a more stable configuration in the potential energy landscape [1,2]. Structural relaxation of silicate glass has been observed when annealing below the glass transition temperature [3-5]. However, it is debatable whether the Kauzmann temperature $T_k$ [6], which is considered the ideal glass transition temperature as no relaxation occurs below $T_k$, exists or not. Although some studies have assumed the existence of $T_k$ [5,7], a report does not agree with the its existence as $S_{conf}(T)$ reaches zero at 0 K [6]. In an experiment [9], a mixed alkali glass was prepared and left at room temperature for 1.5 years, resulting in volume shrinkage. For industrial applications, volume relaxation [3-5], namely, density change, of glasses below $T_g$ is a fundamental issue because the thermal shrinkage of glass occurs at 220 K below the strain point in precise glass manufacturing of high-resolution displays [10]. Although the high thermal stability of glass substrates is necessary for high-resolution displays, the microscopic mechanism of thermal shrinkage is not clear.

Based on the microscopic structure, the thermal shrinkage below the glass transition temperature should be related to the ring size and shape. According to a classical molecular dynamics (MD) study of soda-lime silicate (SLS) glass (composition: $65SiO_2$-$15Na_2O$-$20CaO$ (mol%)), during annealing below $T_g$, Na ions penetrate the six-membered silicate rings, which repairs the acute O-O-O angles of the energetically unstable rings [11]. Another study on sodium silicate glass (composition: $70SiO_2$-$30Na_2O$ (mol%)) suggested that, as the cooling rate decreases, the Si-O-Si bond angle increases, and the opening of the ring can lead to the formation of large pockets of Na ions [12]. However, to the best of our knowledge, no experimental evidence supports these conclusions.

In this study, we demonstrate an increase in the Si-O-Si bond angle during annealing below $T_g$ by measuring the Raman spectra of SLS glass and assigning unknown Raman peaks [13-15]. To assign the peaks, we analyzed the Raman spectra and deconvoluted the structural units using ab initio calculations. Next, we discussed how an increase in the Si-O-Si bond angle contributes to volume shrinkage.

## II. METHODS

### A. Experiment

#### 1. Evaluation of Volume Relaxation

A SLS glass with a composition of $75SiO_2$-$16.7Na_2O$-$8.3CaO$ (mol%) (sample size: $95 \times 8 \times 1$ mm$^3$) was prepared using a conventional melt-quenching method. The glass composition was the same as that used in a previous study, and its $T_g$ and fragility were 804 K and 40, respectively [4]. To erase the thermal history, the sample was annealed at 10 K above $T_g$ ($T_g/T = 0.988$, where $T$ is the annealing temperature) for 2 or 3 h in an electric furnace (KBF748N1, Koyo Thermo Systems Co., Ltd.). Subsequently, the glass sample was annealed at 70 K below $T_g$ ($T_g/T = 1.095$) for 0.5, 2, 6, 8, 24, 73, 168, 264, 380, 672, and 1200 h and immediately quenched to room temperature by being cast on a graphite plate. The change in glass sample length was measured using a microhead spectral interference laser displacement meter (SI-F Series, Keyence). The volume shrinkage rate, $C(t)$, was calculated using the following equation:

$$C(t) = \frac{L(0)^3 - L(t)^3}{L(0)^3} \tag{1}$$

where $L(t)$ is the length of the glass sample and $t$ is the annealing time.

In the glass relaxation glass, the temporal change in the physical properties can be generally described using a Kohlrausch-Williams-Watts (KWW) function: [16]

$$\varphi(t) = \exp\left[-\left(\frac{t}{\tau}\right)^\beta\right] \tag{3}$$

where $\tau$ is the relaxation timescale and $\beta$ is the stretched exponent satisfying $0 < \beta \leq 1$. A lower value of $\beta$ corresponds to a larger distribution of structural relaxation modes. On the other hand, $\varphi(t)$ is expressed using $C(t)$:

$$\varphi(t) = \frac{C(\infty) - C(t)}{C(\infty)} \tag{4}$$

Using equations (3) and (4), we define the fitting function of the volume shrinkage rate $C(t)$ as follows:

$$C(t) = C(\infty)\left(1 - \exp\left[-\left(\frac{t}{\tau}\right)^{\beta}\right]\right) \tag{5}$$

### 2. Raman Spectroscopy

Raman spectroscopy was used to evaluate the temporal changes in the SLS glass structure during annealing below $T_g$. The glass sample for measurement was the same as in Section **1**. All Raman spectroscopy measurements were made at room temperature using confocal Raman microscopy (Nanofinder30, Tokyo Instruments Co.), in which a 532-nm excitation laser beam was focused using a 50× objective lens (NA = 0.80; Nikon Plan Fluor) at a depth of 15 $\mu$m from the surface. Grating was done at 1800 gr/mm corresponding to the wavenumber range of 550–1330 cm$^{-1}$. Additionally, we annealed the sample at 30 K above $T_g$ ($T_g/T = 0.964$) for 2 h, measured the Raman spectra, and compared the results to the one annealed at 10 K above $T_g$, for 2 h.

## B. Simulation methods

### 1. Raman spectra

For comparison with the experimental Raman spectra, density functional theory (DFT) calculations were performed using a glass model comprising 137 atoms (O, Si, Na, and Ca included 82, 35, 16, and 4 atoms, respectively). All DFT calculations were performed using the Quantum ESPRESSO package [17]. The pseudopotential was norm-conserving, and the exchange-correlation function was GGA-PBEsol [18]. The energy cutoff was set to 100 Ry, and a 2 × 2 × 2 k-mesh was used.

First, a glass model was obtained by conducting classical MD simulations using the LAMMPS package [19]. The temperature profile for the fabrication of the glass structures is shown in Fig. 1. After heating the random configurations to 3500 K for 10 ps using the velocity scaling method (NVE ensemble), the glass models were equilibrated at the same temperature for 1000 ps and then quenched to 300 K at a constant cooling rate with an NPT ensemble run. Subsequently, the glass models were equilibrated at 300 K for 3000 ps. Cooling simulations were conducted at three cooling rates— 0.1, 0.01, and 0.001 K/ps—to investigate effects of cooling rate on the Raman spectra and glass microstructures. For the statistical analysis, five independent models were obtained using different initial configurations for each cooling condition.

The classical MD simulations were performed with a Coulomb-Buckingham-type potential:

$$V_{ij} = \frac{1}{4\pi\varepsilon_0}\frac{z_i z_j e^2}{r} + A_{ij}\exp\left(-\frac{r}{B_{ij}}\right) - \frac{C_{ij}}{r^6}, \tag{6}$$

where $z$ is the effective ionic charge, $r$ is the distance between two ions, and $\varepsilon_0$ is dielectric constant of vacuum. $A_{ij}$, $B_{ij}$ and $C_{ij}$ are the optimized parameters for each ion pair [20]. The first, second, and third terms represent the Coulombic interaction, short-range repulsive force, and short-range attractive force, respectively. The short-range cutoff distance was 6 Å. The particle-particle mesh Ewald (PPPM) method [21] was used to evaluate the Coulomb interactions. The simulation box was a cubic cell, and periodic boundary conditions were applied in all directions. A time step of 1.0 fs was used to integrate the atomic motions. Larger glass models comprising 18001 ions were also constructed by applying four different cooling rates (10, 1, 0.1, and 0.01 K/ps). The other calculation conditions were consistent with those of the small model, except for the cutoff distance, which was set to 10 Å for the larger models.

After classical MD simulations, *ab initio* MD (AIMD) simulations were conducted at 0 K to refine short-range structures, such as chemical bond lengths and angles. Similar to previous studies [22,23], the medium-range structures represented by $Q_n$ units and ring size distributions were not altered by the AIMD simulations. Structural relaxation calculations were performed using DFT to evaluate the vibrational eigenmodes, frequencies, and Raman tensors of the glass models. Following a previous study [22], the discrete spectra were broadened by using Gaussians with a 20 cm$^{-1}$ FWHM. All the Raman spectra of the five independent models were averaged and normalized for comparison with the experimental spectra. The experimental data were corrected by multiplying the intensity factor by the following equation [22] for appropriate comparison with the theoretical Raman spectra:

$$\frac{\omega}{n(\omega)+1}, \quad n(\omega) = \frac{1}{\left(\exp\left(\frac{\omega}{k_B T}\right) - 1\right)} \tag{7}$$

where $n(\omega)$ is the Bose factor and $k_B$ is Boltzmann constant.

## 2. Structure analysis

After *ab initio* relaxation simulations, the distributions of the Si-O-Si angle, $Q_n$, $Q_n$-O-$Q_m$ angle, ring size, and O-O-O angle in the ring structures were analyzed. The ring size represents the number of Si atoms in a ring structure and was calculated using an algorithm exploring the shortest path proposed by King et al. [24] Dijkstra's method was used to calculate the shortest path [25].

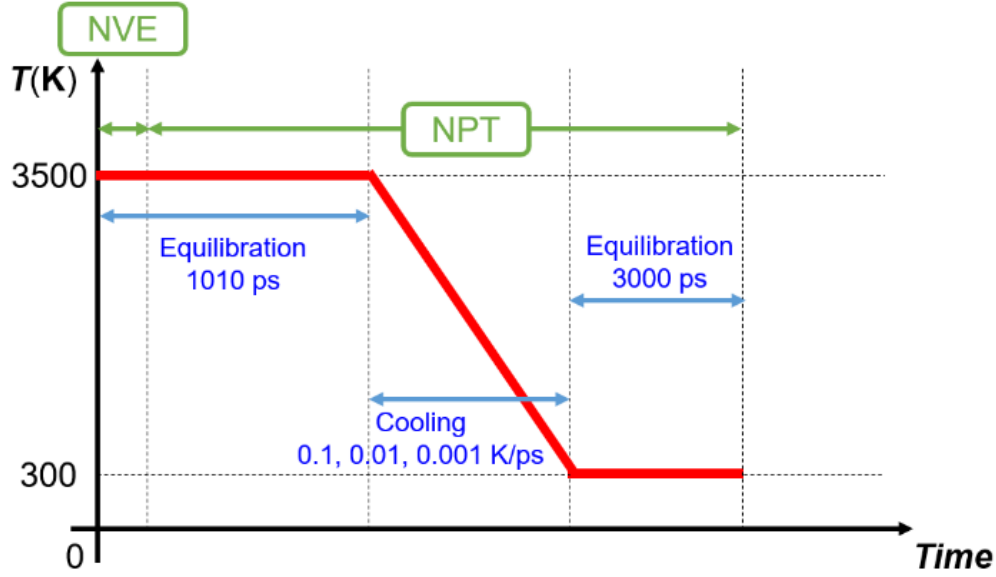

FIG. 1. Temperature profile used to obtain SLS glass at 300 K in classical MD simulation.

### 3. Deconvolution of Raman spectra

To identify the structural origin of each Raman peak in the high-frequency region, the theoretically calculated Raman spectra were decomposed as follows: The component $R_{ij}^p$ of the Raman susceptibility tensor, which is a second-order symmetric tensor in Raman scattering, was defined by the following equation:[22,26,27]

$$R_{ij}^p = \sqrt{V} \sum_{I=1}^{N} \sum_{k=1}^{3} \frac{\partial x_{ij}}{\partial R_{I,k}} \frac{e_{I,k}(\omega_p)}{\sqrt{M_I}} \qquad (8)$$

where $V$ is the volume of the system, $N$ is the number of atoms, and $M_I$ is the mass of atom $I$. $\frac{\partial x_{ij}}{\partial R_{I,k}}$ and $e_{I,k}(\omega_p)$ are the 3 × 3 Raman tensor and the 3-component of the phonon eigenvectors of atom $I$ in the $3N$ eigenfrequencies $\omega_p$, respectively. Here the Raman tensor $\frac{\partial x_{ij}}{\partial R_{I,k}}$ is given by (9) below.

$$\frac{\partial x_{ij}}{\partial R_{I,k}} = \frac{1}{V} \frac{\partial^2 F_{I,k}}{\partial \varepsilon_i \varepsilon_j} \qquad (9)$$

where $F_{I,k}$ is the force in the applied electric field $\varepsilon$ of atom $I$.

The entire Raman susceptibility tensor can be decomposed into contributions of several specific groups, $\gamma$, as,

$$R_{ij}^{p,\gamma} = \sqrt{V} \sum_{I}^{\gamma} \sum_{k=1}^{3} \frac{\partial x_{ij}}{\partial R_{I,k}} \frac{e_{I,k}(\omega_p)}{\sqrt{M_I}} \qquad (10)$$

The atom types, $\gamma$, were defined by (i) elements, (ii) bridging oxygen (BO) or non-bridging oxygen (NBO), (iii) Si-O-Si angle, and (iv) $Q_n$-O-$Q_n$ angle to determine their contributions to an entire Raman spectrum. Here, it should be noted that, as discussed in D. Kilymis, et al [22], the sum of the decomposed Raman spectra does not necessarily equate to the original Raman spectrum. Therefore, the minimal differences between the decomposed and original Raman spectra were acceptable.

From the components of the Raman susceptibility tensor, the trace of susceptibility, $\alpha_p$, and anisotropy of susceptibility, $\tau_p$, are derived as

$$\alpha_p = \frac{R^p_{11} + R^p_{22} + R^p_{33}}{3} \tag{11}$$

$$\tau_p^2 = \frac{\left(R^p_{11} - R^p_{22}\right)^2 + \left(R^p_{22} - R^p_{33}\right)^2 + \left(R^p_{33} - R^p_{11}\right)^2 + 6\left({R^p_{12}}^2 + {R^p_{13}}^2 + {R^p_{23}}^2\right)}{2} \tag{12}$$

Consequently, the Raman activity $I_p$ of a vibration mode $p$ is obtained.

$$I_p = 45{\alpha_p}^2 + 7\tau_p^2 \tag{13}$$

### III. RESULTS

#### A. Experimental results

##### 1. Volume relaxation annealing below $T_g$

The shrinkage rate and relaxation function obtained by the volume relaxation measurements at $T_g$-70 K ($T_g/T = 1.095$) are shown in Figs. 2 (a) and (b), respectively. According to equation (5), $C(\infty)$, $\beta$, and $\tau$ were estimated to be 3950±156 ppm, 0.32±0.02, and 58.5±15.5 h, respectively. Compared with the results of the previous study [4] at $T_g$- 50 K annealing, $C(\infty)$ is larger, $\beta$ is smaller, and $\tau$ is larger. This is because heat treatment at lower temperatures is expected to cause more shrinkage and require a longer time owing to the presence of more volume relaxation modes.

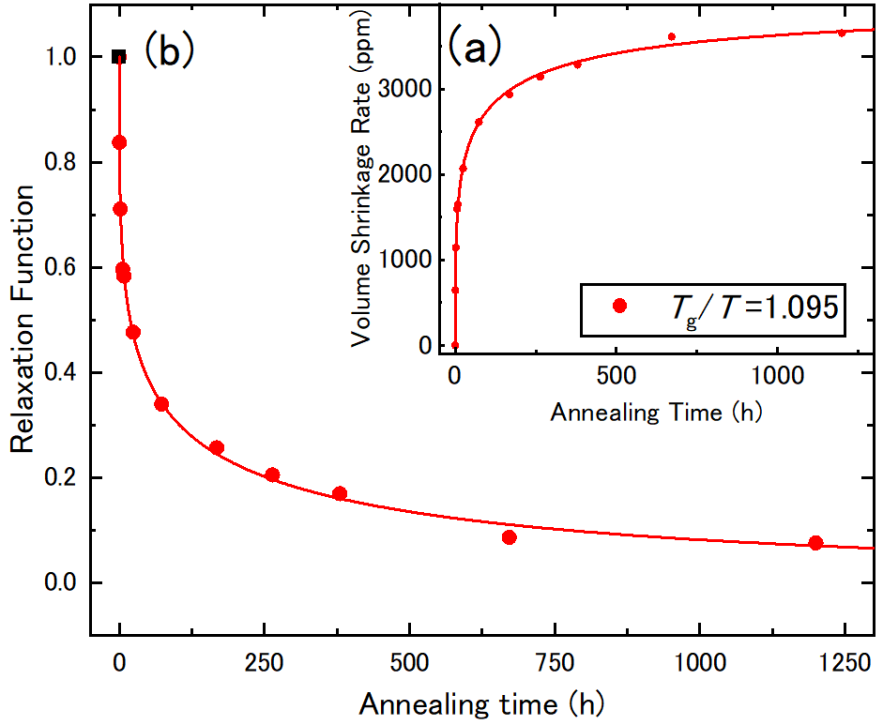

FIG. 2. (a) Volume shrinkage rate, and (b) relaxation function at different annealing times. Lines represent the fitting results using equations (5) and (3).

### 2. Experimental Raman spectra

The Raman spectra of the SLS glass samples annealed at $T_g$ = 70 K for 0 to 1200 h are shown in Fig. 3, and the extended figures at 800, 950, 1050, and 1100 cm$^{-1}$ are illustrated in Figs. 4 (a), (b), (c), and (d), respectively. Consequently, it can be observed that the peak intensity at 800 cm$^{-1}$ decreases even in the late stage of structural relaxation (Fig. 4 (a)). The band at 800 cm$^{-1}$ is attributed to both the motion of Si atoms against the tetrahedral oxygen cage [3] and the bending mode of the Si-O-Si bridge [22]. In addition, the Raman spectra at approximately 1050 and 1100 cm$^{-1}$ shifted toward higher wavenumbers, even during the later stages of volume relaxation (Figs. 4 (c), (d)). Conversely, the intensity at 950 cm$^{-1}$ decreased, especially in the early stages of structural relaxation (Fig. 4 (b)).

These high-frequency bands (from 900 to 1300 cm$^{-1}$) correspond to the stretching and bending vibrations of the $Q_n$ structures, as discussed in previous studies [3,13-15]. The peak at 950 cm$^{-1}$ corresponds to the Si-O stretching vibration mode in the $Q_2$ and $Q_3$ structures [3,28-31], whereas that at 1100 cm$^{-1}$ is attributed to the antisymmetric stretching vibration mode of BO comprising the $Q_4$ structure [3,28-31]. In contrast, the structural origin of the peak at 1050 cm$^{-1}$ has yet to be clarified [3,13-15]. Therefore, the peak at 1050 cm$^{-1}$ is treated as an unassigned peak, and the variation of the peak area with annealing time will be discussed.

The experimental Raman spectra observed from 900 to 1300 cm$^{-1}$ were decomposed by four Gaussian bands, and each peak area, $A_i$, at $i$ cm$^{-1}$ was evaluated. The peak positions at 1050 cm$^{-1}$ and 1150 cm$^{-1}$ were fixed to prevent the peak area underestimation at 1100 cm$^{-1}$ during fitting. Here, these peak positions are arbitrarily defined, whereas index $i$ is denoted as 800, 950, 1050, and 1100 by considering the peak positions for convenience. The variations in the ratios of $A_{950}/A_{1100}$, which correspond to $Q_2/Q_3$, and $A_{1050}/A_{1100}$ with annealing time, are shown in Figs. 5(a) and (b), respectively. The rapid decrease of $A_{950}/A_{1100}$ and its saturation at approximately 672 h suggests that the $Q_2$ species was converted into $Q_3$ over time by the reaction of $Q_2+Q_4 \rightarrow 2Q_3$ at the early stage of structural relaxation. In contrast, $A_{1050}/A_{1100}$ decreased even in the later stages until 672 h. The shift to a higher wavenumber at 1100 cm$^{-1}$ (Fig. 4(d)) corresponds to a decrease in $A_{1050}$. The reason behind this is discussed in Section III. The Raman spectra of the glass annealing at $T_g+30$ K ($T_g/T = 0.964$) for 2 h, $T_g+10$ K for 2 h, and $T_g-70$ K ($T_g/T = 1.095$) for 1200 h are shown in Fig. S1, and the extended figures of these spectra at 950, 1050, and 1100 cm$^{-1}$ are shown in Fig. S2. Based on these results, we conclude that when SLS glass is heat treated at temperatures below $T_g$, (i) the reaction $Q_2+Q_4 \rightarrow 2Q_3$ proceeds in the early stage, and (ii) $A_{1050}$ decreases even in the later stages of structural relaxation.

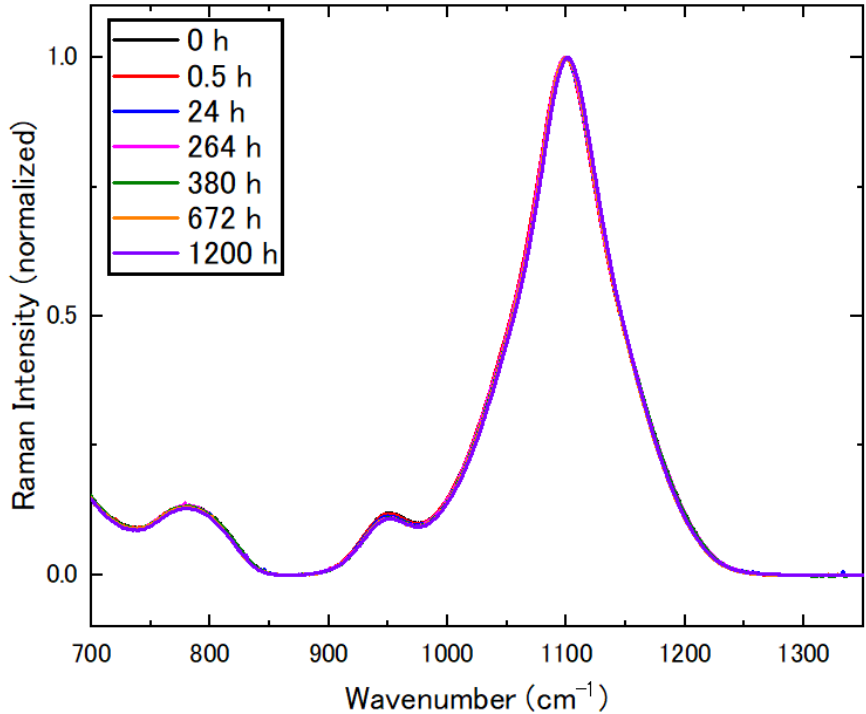

FIG. 3. Raman spectra of different annealing times at $T_g-70$ K.

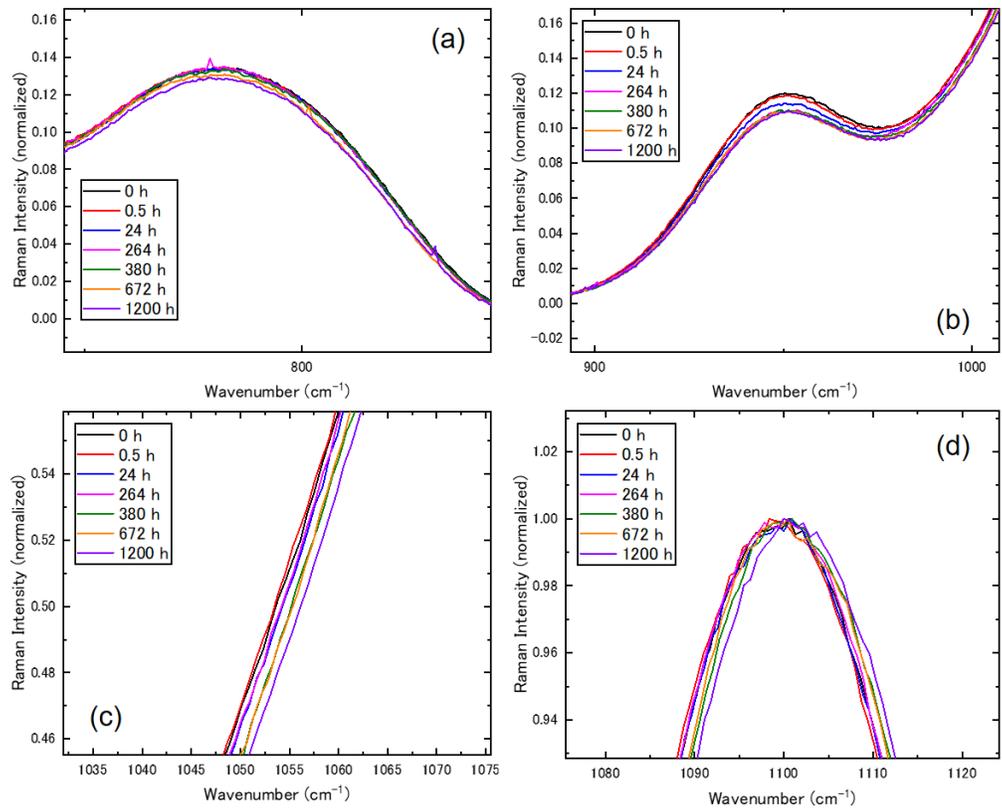

FIG. 4. Extended figures of Fig. 3. for the wavenumbers of (a)800, (b)950, (c)1050, and (d)1100 cm$^{-1}$.

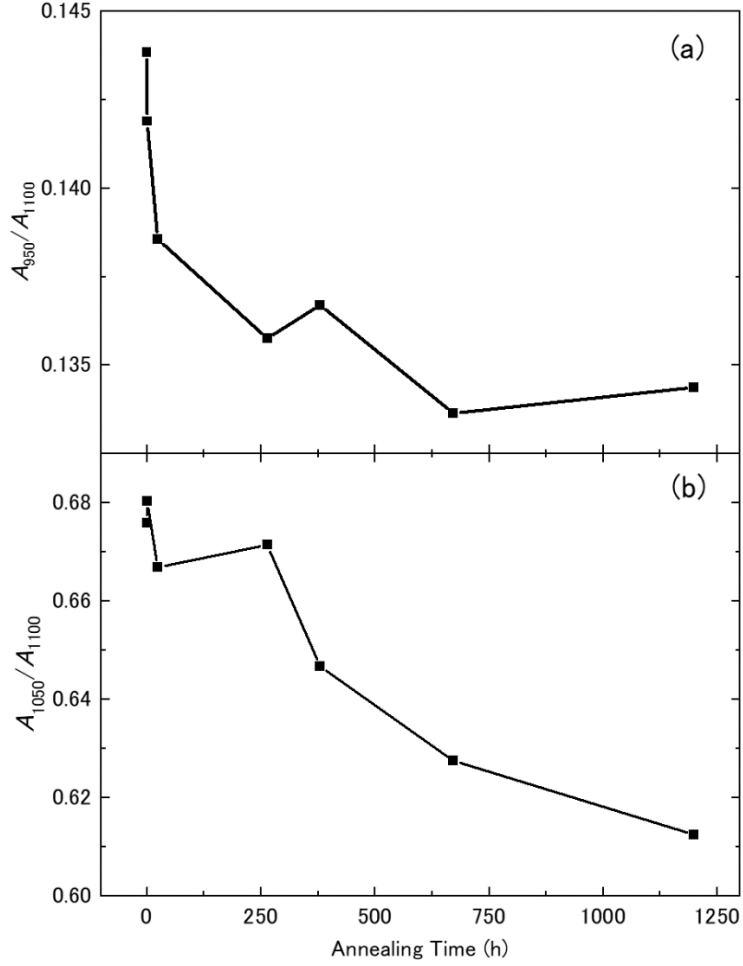

FIG. 5. Peak area ratios of (a) $A_{950}/A_{1100}$ and (b) $A_{1050}/A_{1100}$ with annealing time.

### B. Calculation results

#### 1. Comparison of experimental and calculated Raman spectra

The Raman spectra computed for the three models obtained at three different cooling rates were compared with the experimental Raman spectra after annealing in Fig. 6. Notably, the main peaks of the calculated spectra were downshifted by approximately 50–100 cm$^{-1}$ compared to the experimental spectra, analogous to previous DFT studies [32-34]. This discrepancy may be due to the limited accuracy of the GGA-PBEsol method in estimating the bonding energy.

The glass model constructed at a slower cooling rate is expected to exhibit a lower fictive temperature, as shown in a previous study that examined the cooling rate dependence on density and structure [12]. Similar to the experimental Raman spectra, the theoretical spectra show a decrease in the peak height at approximately 900–950 cm$^{-1}$, which corresponds to the $Q_2$ species. Moreover, a decrease in the peak intensity of the 1100–1200 cm$^{-1}$ range, which is attributed to the $Q_4$ species, was observed at a slower cooling rate. The narrower width of the peak ranging from 1000 to 1100 cm$^{-}$ [1] for the models obtained at slower quenching might be attributed to a more ordered microstructure with a lower fictive temperature, as discussed in the structural analysis that follows.

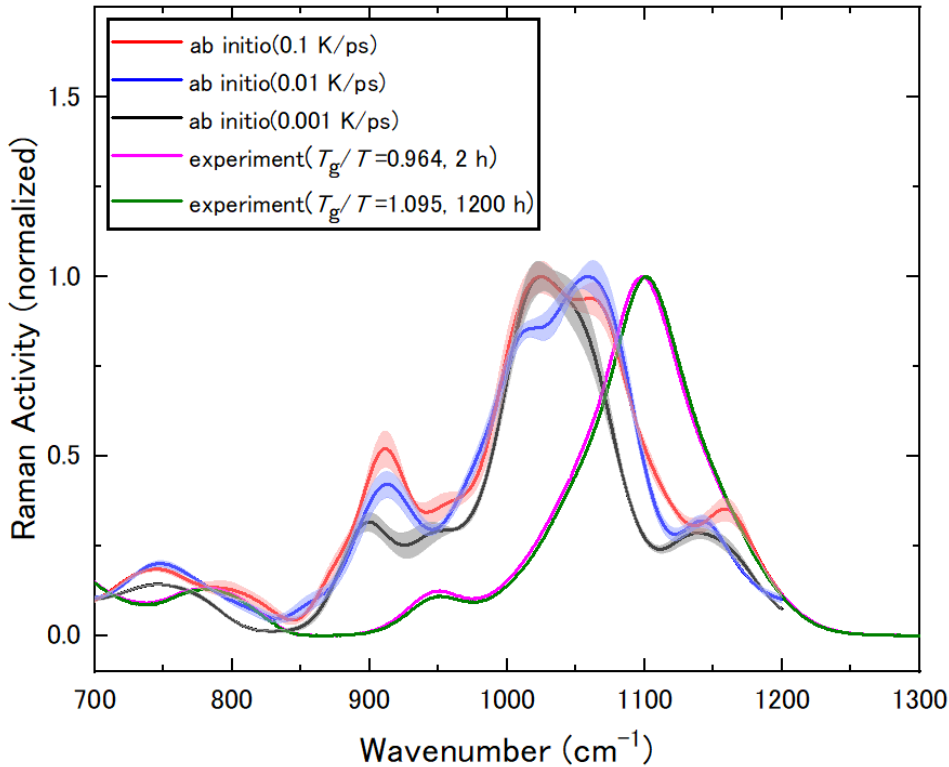

FIG. 6. Calculated Raman spectra for models obtained with three different cooling rates together with the experimental Raman spectra of the samples annealed at $T_g + 30$ K for 2 h and $T_g - 70$ K for 1200 h. Filled areas represent a standard deviation of 5 runs at each point.

### 2. Variations in medium-range structure

To understand the effects of the cooling rate on the $Q_n$ speciation, bond angle, and ring size distributions, structural analyses were performed. The $Q_n$ and $Q_n$-$Q_m$ distributions obtained from the 137- and 18001-ions systems are listed in Table. I. According to the $Q_n$ distribution, $Q_3$ increases monotonically, whereas $Q_1$, $Q_2$, and $Q_4$ decrease at a slower cooling rate. $Q_1$ structures exist because of the much faster cooling rate in MD simulations, although MAS NMR studies have rarely observed unstable structural units [35,36]. As in the experiments done by previous studies [3,12], the slower cooling promoted the reaction $Q_2+Q_4 \rightarrow 2Q_3$.

According to the $Q_n$-O-$Q_m$ distribution, the $Q_4$-$Q_2$ and $Q_4$-$Q_1$ structures, which might exhibit higher entropy, decreased with decreasing cooling rates. Conversely, the $Q_3$-$Q_3$ structure with lower entropy increased. These structural variations suggest that a more ordered network structure with energetically stable $Q_3$ units was developed by a longer relaxation.

Figure 7 shows the Si-O-Si angle distributions of the two systems. A decrease in the small Si-O-Si angles was observed in both systems with lower cooling rates, as in previous studies [12,37]. Indeed, the $Q_n$-O-$Q_m$ units with more BOs ($Q_{n,m}$ with larger $n$ and $m$) exhibited a wider Si-O-Si angle, as shown in Fig. S3. This observation is consistent with a previous study [22], which indicated that slower cooling increases $Q_3$ and extends the silicate network. Therefore, the reduction in structures with small Si-O-Si angles may facilitate the incorporation of Na ions into the ring structures.

Based on the ring size distributions illustrated in Fig. 8, the presence of six-membered ring dominates the SLS glass. Notably, in the smaller model, four- and five-membered rings decreased with a decreasing cooling rate. Similarly, three- and four-membered rings decreased in the larger system. These results imply that longer relaxation enhances the stability of the glass microstructure by eliminating rings with less than six members, which are topologically overconstrained and thus exhibit significant internal stress [37]. To prove this hypothesis, the O-O-O angles in the rings were analyzed, as illustrated in Fig. 9, because acute O-O-O angles reduce the space around the ring center, as discussed in M. Shimizu, et al [11]. The number of acute O-O-O angles decreased in the five-, six-, and seven-membered rings, indicating that the rings repair their distorted structures by longer relaxation. In summary, the slower cooling promoted the reaction $Q_2+Q_4 \rightarrow 2Q_3$, and the resulting increased $Q_3$ widened the ring structures.

TABLE. I. $Q_n$ and $Q_n$-$Q_m$ distributions of 137- and 18001-ion systems at different cooling rates. The values were added for five runs and the standard deviation was noted.

|  | 137 ions system (*ab initio* MD) | | | 18001 ions system (classical MD) | | | |
| --- | --- | --- | --- | --- | --- | --- | --- |
|  | 0.1 K/ps | 0.01 K/ps | 0.001 K/ps | 10 K/ps | 1 K/ps | 0.1 K/ps | 0.01 K/ps |
| $Q_4$ | 82 ± 1.14 | 77 ± 1.34 | 73 ± 0.89 | 2143 | 2135 | 2114 | 2124 |
| $Q_3$ | 69 ± 1.92 | 78 ± 2.51 | 84 ± 1.79 | 1936 | 1956 | 1988 | 1986 |
| $Q_2$ | 21 ± 0.83 | 18 ± 1.34 | 18 ± 0.89 | 504 | 481 | 489 | 459 |
| $Q_1$ | 3 ± 0.55 | 2 ± 0.55 | 0 ± 0 | 42 | 53 | 36 | 57 |
| $Q_4$-$Q_4$ | 81 ± 3.03 | 79 ± 1.92 | 70 ± 2.35 | 2401 | 2347 | 2328 | 2354 |
| $Q_4$-$Q_3$ | 136 ± 3.42 | 130 ± 1.58 | 134 ± 2.49 | 3192 | 3278 | 3278 | 3273 |
| $Q_4$-$Q_2$ | 28 ± 1.87 | 19 ± 3.19 | 18 ± 1.95 | 545 | 534 | 502 | 488 |
| $Q_4$-$Q_1$ | 2 ± 0.55 | 1 ± 0.45 | 0 ± 0 | 25 | 23 | 20 | 25 |
| $Q_3$-$Q_3$ | 29 ± 2.86 | 46 ± 3.42 | 51 ± 1.79 | 1113 | 1100 | 1140 | 1142 |
| $Q_3$-$Q_2$ | 12 ± 1.22 | 11 ± 1.10 | 16 ± 1.10 | 381 | 361 | 390 | 372 |
| $Q_3$-$Q_1$ | 1 ± 0.45 | 1 ± 0.45 | 0 ± 0 | 7 | 26 | 16 | 27 |
| $Q_2$-$Q_2$ | 1 ± 0.45 | 3 ± 0.55 | 1 ± 0.45 | 36 | 31 | 43 | 26 |
| $Q_2$-$Q_1$ | 0 ± 0 | 0 ± 0 | 0 ± 0 | 10 | 4 | 0 | 5 |

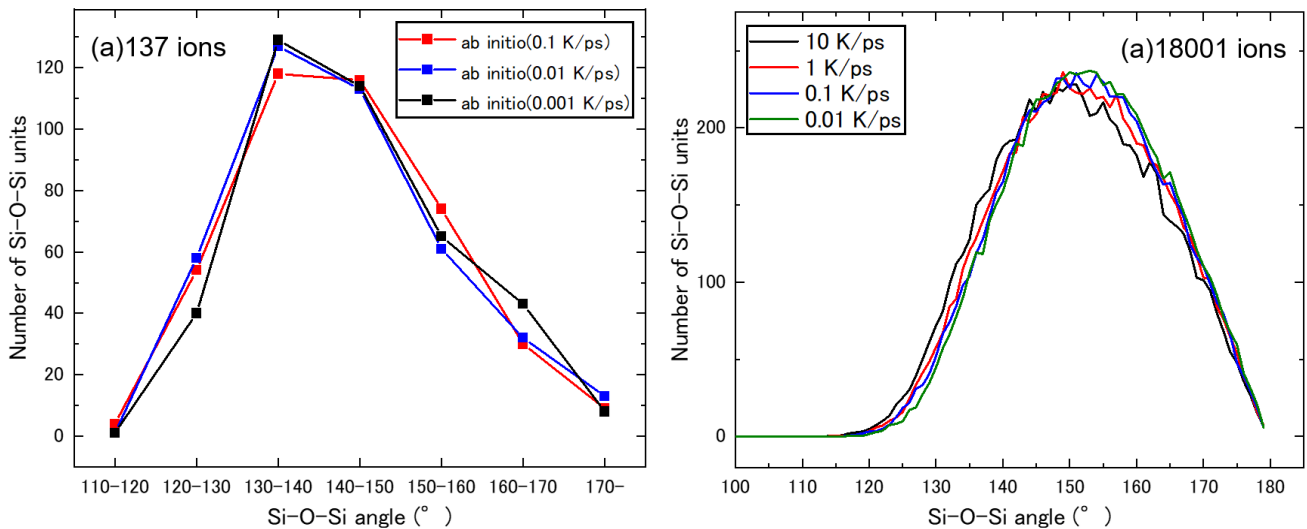

FIG. 7. Si-O-Si angle distribution of (a) 137-ion system after ab initio relaxation and (b) 18001-ion system after classical MD.

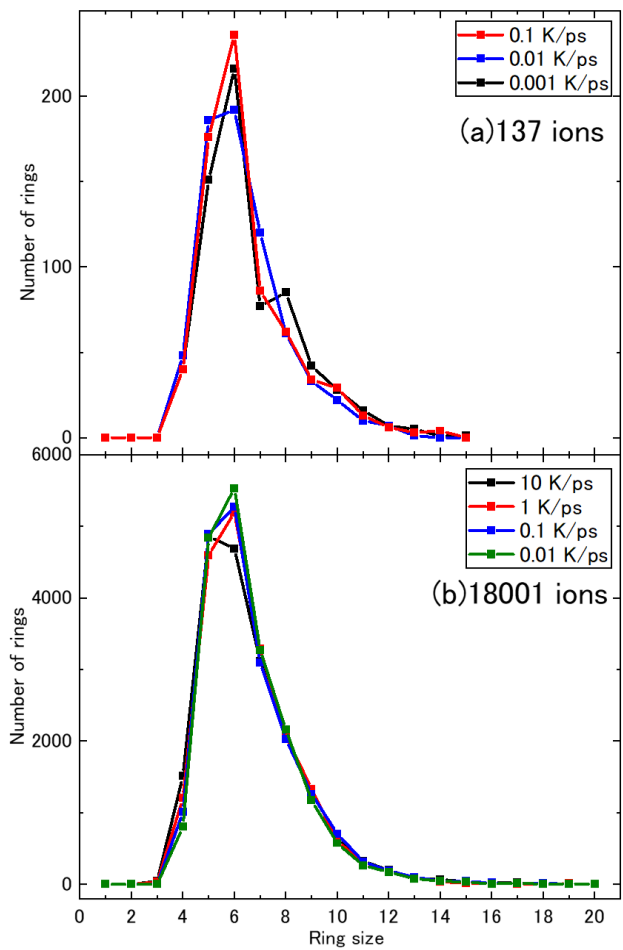

FIG. 8. Ring size distribution of (a)137-ion system after *ab initio* relaxation and (b)18001-ion system after classical MD.

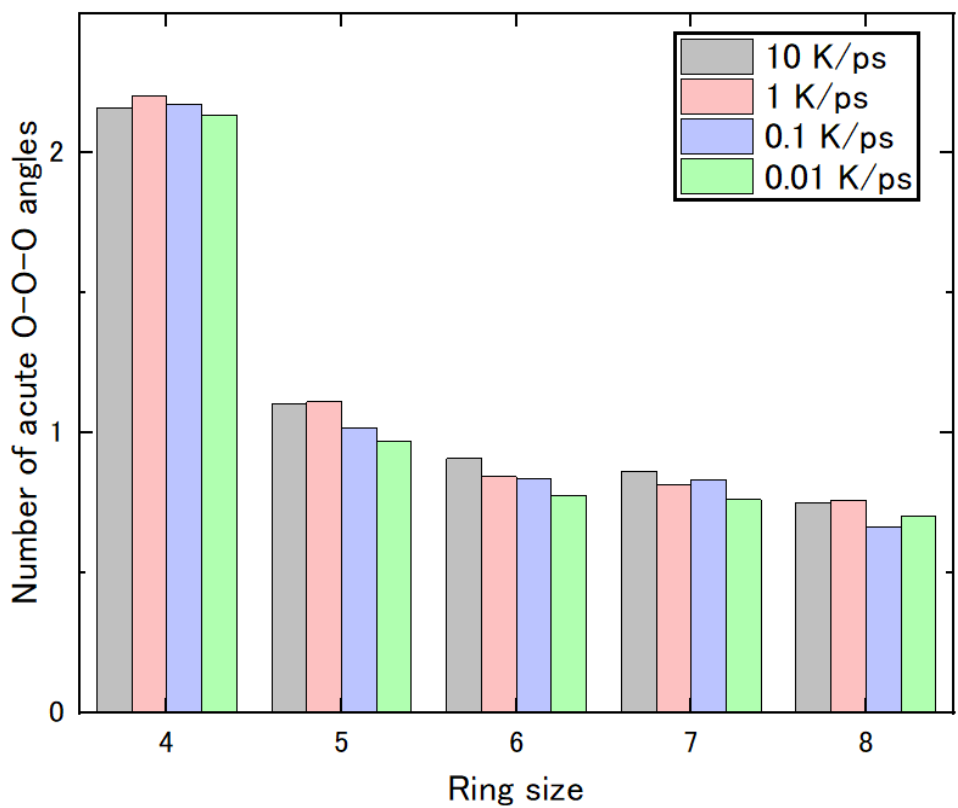

FIG. 9. Number of acute O-O-O angles in silicate rings.

### 3. *Decomposition of calculated Raman spectra*

To obtain further insight into the relationship between the microstructures and Raman spectra, the theoretical spectra were decomposed by considering the element types, as shown in Fig. 10(a). The models obtained at the three cooling rates are compared in Figure 10(b).

Below 400 cm$^{-1}$, the network modifier cations Na and Ca dominate Raman scattering, as discussed in previous studies [22,38,39]. Notably, the Raman activity of Na increased, whereas that of Ca decreased monotonically with decreasing cooling rates. It can be inferred that these results reflect changes in the coordination environment of the modifier cations. The details are discussed in Section IV.

In the high-frequency region ranging from 800 to 1200 cm$^{-1}$, the decomposed Raman spectra of O and Si varied with cooling rate. Si had the highest intensity at approximately 750 cm$^{-1}$, which corresponds to the motion of Si atoms against O atoms [3]. Because the intensity of the band at 750 cm$^{-1}$ of Si ions did not change with the cooling rate, the variation in the intensity at 800 cm$^{-1}$ in the experimental Raman spectra is not related to the motion of the Si ions.

Because oxygen has the largest contribution to the entire spectrum in any case, the oxygen atoms were decomposed into BO and NBO, and the NBO was further distinguished by the $Q_n$ species bonded to the NBO (denoted as $Q_n$-NBO, hereafter), as illustrated in Fig. 11. Consequently, it was revealed that NBO contributes to Raman scattering at frequencies higher than 750 cm$^{-1}$ compared to BO. Moreover, each $Q_n$-NBO exhibited a different scattering profile. For instance, $Q_3$-NBO exhibited higher Raman scattering at a higher frequency than $Q_2$-NBO, which is consistent with previous experimental Raman spectroscopic studies [3,13-15,28-31]. This is because the Si-O⁻ non-bridging bond becomes stronger as the silicate anion polymerizes, as mentioned in a previous study [31].

Figure 11(b) shows the effect of cooling rate on the decomposed Raman spectra. The high-frequency peak areas of $Q_2$-NBO and $Q_3$-NBO were evaluated using the trapezoidal integration method (Fig. S4). With decreasing cooling rates, the peak area of $Q_2$-NBO decreased, whereas that of $Q_3$-NBO increased. This result implies that the reaction, $Q_2+Q_4\rightarrow 2Q_3$ occurred in the models obtained by the MD simulations with slower cooling, which is consistent with our experimental observations in Fig. 5.

The small peak at approximately 750 cm$^{-1}$ in the decomposed Raman scattering of BO decreased with a slower cooling rate, which might be related to the variation in the experimental Raman spectra at approximately 800 cm$^{-1}$ shown in Fig. 4(a). The other broad peak appeared in the wide frequency range of 900 to 1200 cm$^{-1}$ for BO. Its intensity decreased as the cooling rate decreased, corresponding to a reduction in $A_{1050}$ (Figs. 4 (c) and 5 (b)).

To identify the structural origin of the change above, the BO atoms were further classified based on the Si-O-Si angle at intervals of 10°, as illustrated in Fig. 12. At any cooling rate, the BO of the larger Si-O-Si angles exhibited a wide peak at higher frequencies, while a primary peak at approximately 900–1100 cm$^{-1}$ was attributed to BO in the small Si-O-Si angle interval of 130˚–140˚. Therefore, the unassigned peak at 1050 cm$^{-1}$ in the experimental Raman spectra should be attributed to BO forming small Si-O-Si angles. Because the small angles were ameliorated by the longer cooling procedure, as mentioned previously, the broad peak at approximately 1050 cm$^{-1}$ decreased as the cooling rate decreased in both our experimental and theoretical observations. BOs with small Si-O-Si angles formed bonds with $Q_n$ species with more NBOs, according to Fig. S3. The broad peak was attributed to the stretching and bending vibrational modes, as shown in Fig. S5(a).

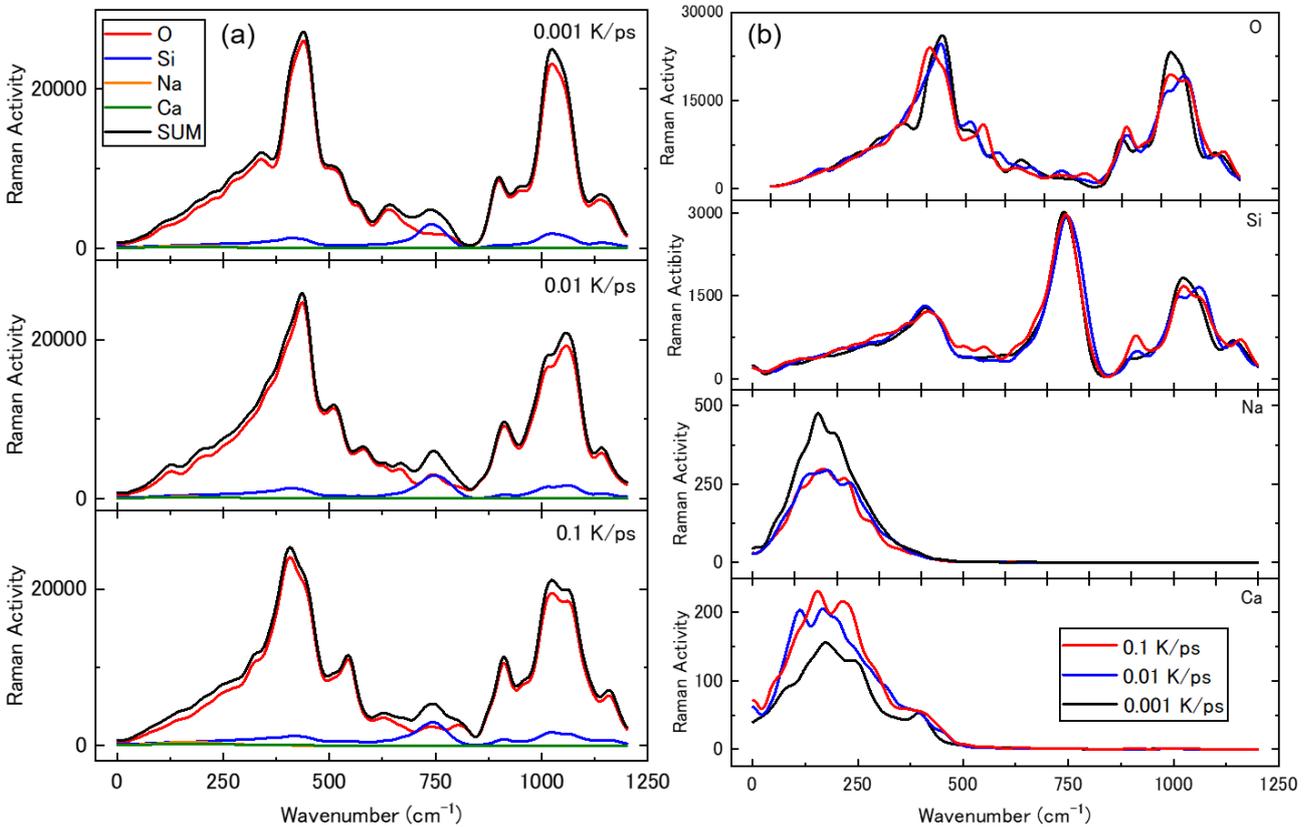

FIG. 10. Comparison of decomposed Raman spectra focusing on (a) each element and (b) different cooling rates.

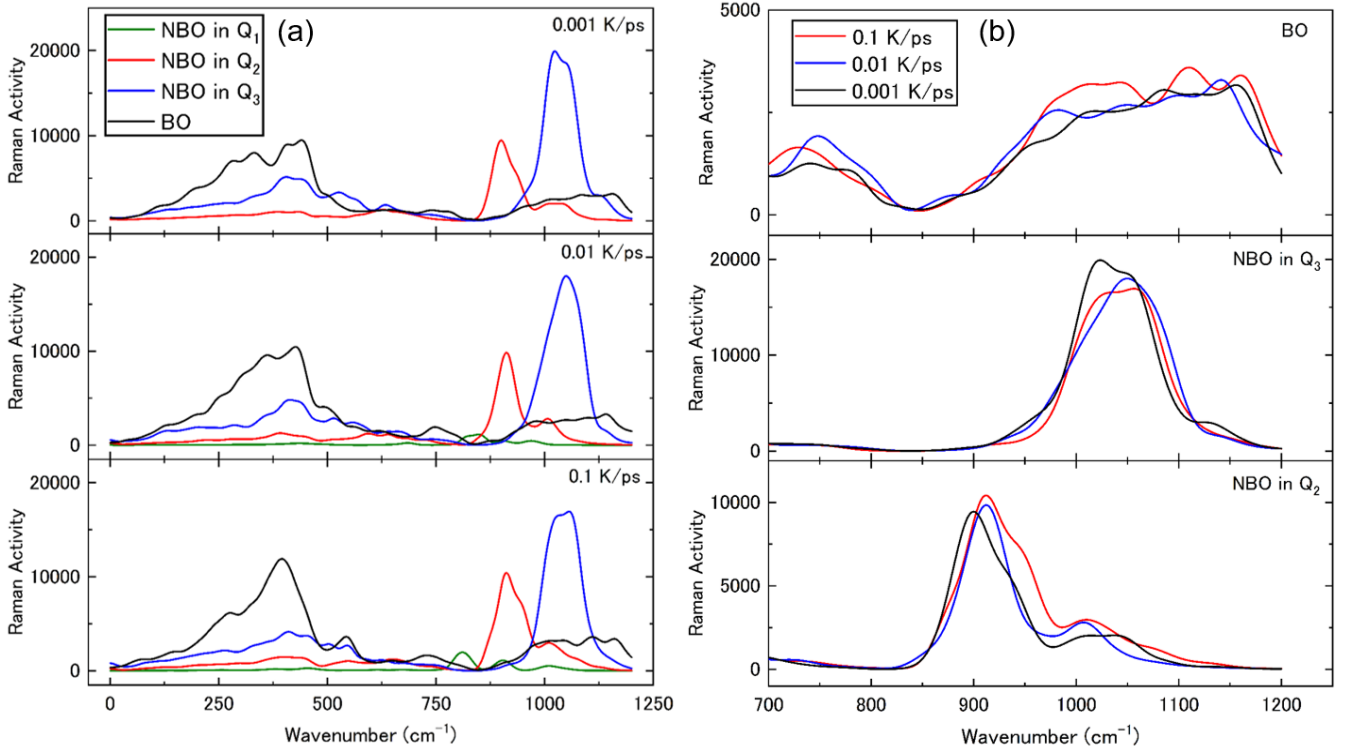

FIG. 11. (a) Decomposed Raman spectra of NBOs and BOs. It was also decomposed by the $Q_n$ species bonded to the NBO. (b) The cooling rate dependence of the decomposed Raman spectra of NBOs and BOs for high-frequency $Q_n$ band.

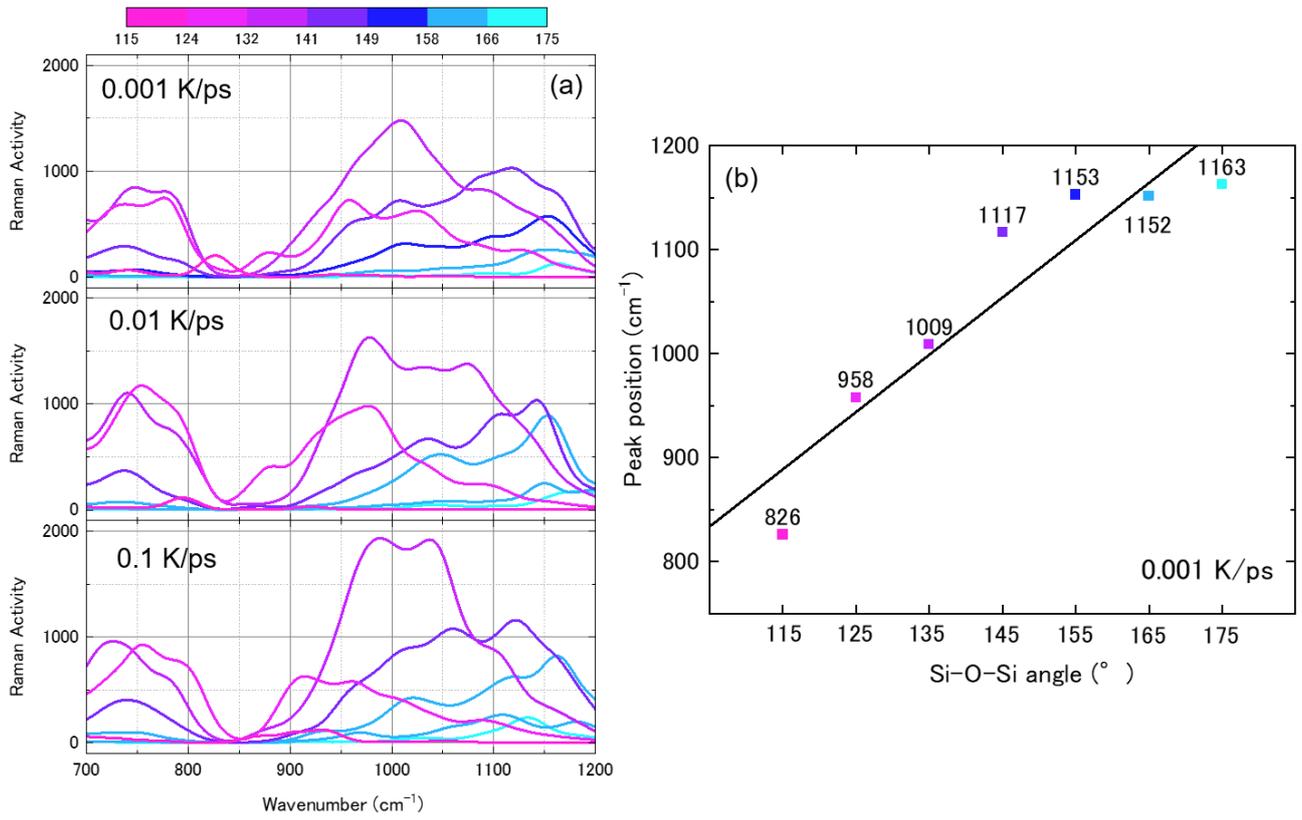

FIG. 12. (a) Si-O-Si angle dependence of decomposed Raman spectra of BOs for each cooling rate. Colors correspond to the size of Si-O-Si bond angle. (b) Plots of peak position with maximum intensity in decomposed Raman spectra at 0.001 K/ps. The colors correspond to those in (a), and the horizontal axis represents the class value.

## IV. DISCUSSION

### A. Mechanism of structural relaxation

Based on the experimental and theoretical Raman spectra, the microscopic mechanisms of structural relaxation in the SLS glass were determined. The decrease in $A_{950}/A_{1100}$ (i.e., $Q_2/Q_3$) in the experimental Raman spectra (Fig. 5(a)) suggested that $Q_2+Q_4 \rightarrow 2Q_3$ is dominant in the early stages of structural relaxation because $A_{950}/A_{1100}$ does not almost change over an annealing period of 672 to 1200 h. Thus, this reaction was dominant during the early stages of glass relaxation. Conversely, a decrease in $A_{1050}/A_{1100}$ (Fig. 5(b)), and the corresponding peak shifts were evident even after 672 h, suggesting that the structural relaxation due to the widening of the Si-O-Si angles became more apparent in the later stage of relaxation. In other words, the reduction in the small Si-O-Si angles while maintaining the $Q_n$ structure leads to slow relaxation in the SLS glass.

The widening of the Si-O-Si angles is attributed to the reduction of three and four-membered rings, as shown in Fig. 13(a), whereas there is another possibility, that is, a change in the ring shapes. To verify this hypothesis, the number of acute O-O-O angles in the six-membered rings, which dominate the ring size distribution, was evaluated for the glass models obtained at cooling rates of 10, 1, 0.1, and 0.01 K/ps, as illustrated in Fig. 14. The relationship between O-O-O angle and Si-O-Si angle in 6-membered ring is also investigated in Fig. S8. It is suggested that small Si-O-Si angle is attributed to small O-O-O angle. Consequently, the average Si-O-Si angles increased as the acute O-O-O angles decreased, which implies that both the reduction of small rings and the widening of the ring shape contributed to slow relaxation in SLS glass.

### B. Effect of decreasing small Si－O－Si angle on volume relaxation

As suggested in X. Li, et al [12], a large Si-O-Si angle is expected in an alkali ion with a pocket-like structure. To verify this, the relationship between the number of acute O-O-O angles in the six-membered ring and the distance between the nearest Na ion and the ring center ($d_{Na-V}$) was investigated, as shown in Fig. 15(a). Consequently, it was revealed that the Na ions were located closer to the six-membered ring with less acute O-O-O angles. This result is consistent with that reported by M. Shimizu, et al [11], which investigated the structural relaxation of SLS glass using microsecond timescale MD simulations.

Next, the effect of these microstructure rearrangements on volume reduction was investigated. A Voronoi volume analysis was performed for a system comprising 18001 ions using the Voro++ package [40], assuming that the ionic radii of O, Si, Na, and Ca are 1.38, 0.26, 1.02, and 1.0 Å, respectively. The Voronoi volume is defined as the volume of the Voronoi cell given by the polyhedron around the center of an ion, consisting of perpendicular bisecting planes between the central ion and its neighboring ions. Fig. 15(b) illustrates the relationship between $d_{Na-V}$ and Voronoi volumes of the six-membered ring with and without the nearest Na ion. Consequently, six-membered rings with a Na ion within a shorter range exhibited smaller volumes. Therefore, it can be inferred that the slow relaxation caused by the widening of the Si-O-Si angle allowed the Na ion to come closer to the six-membered rings, thereby reducing the volume locally. Conversely, the number of three- and four-membered rings, decreased, as shown in Fig. 13(b), which might result in an increase in the local volume. These two competitive structural variations may result in subtle variations in the whole volume, as observed in our experiments.

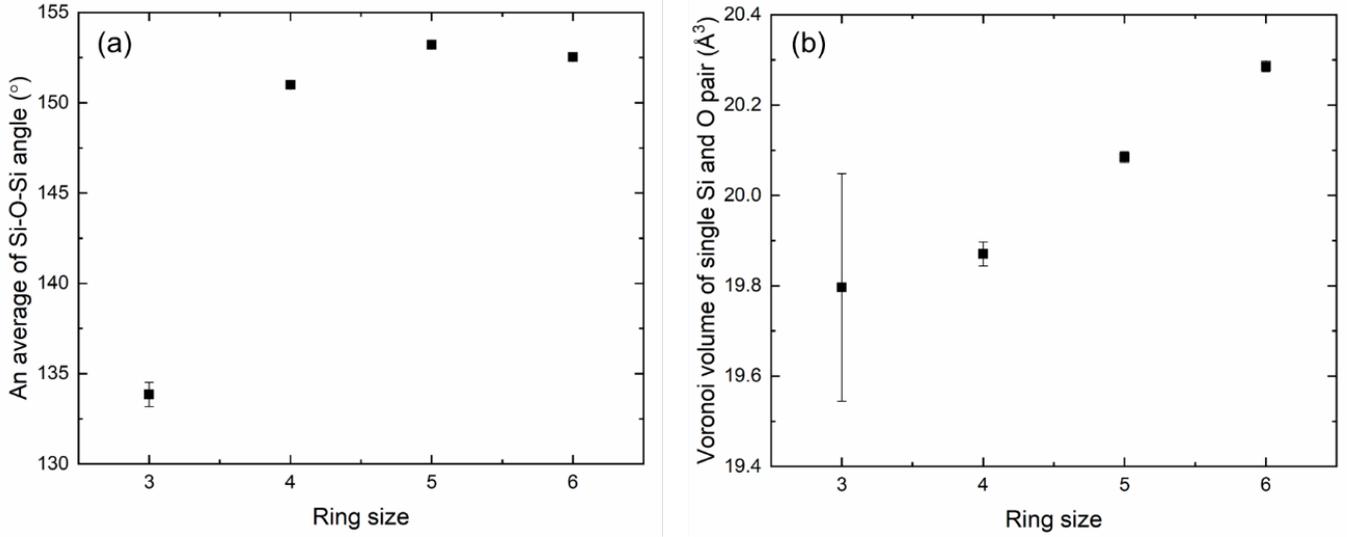

FIG. 13. Ring-size dependence of (a) Si-O-Si angle and (b) Voronoi volume of single Si and O pair. The data presents the summary of the glasses obtained by cooling rates of 10, 1, 0.1, and 0.01 K/ps. To calculate the Voronoi volume of a single Si and O pair, we calculated the Voronoi volume of each n-membered ring and divided it by n to determine the effect of ring number change on the molar volume.

### C. Position change of network modifier cations under volume relaxation

As shown in Fig. 10, Na and Ca exhibited opposite intensity changes in the decomposed Raman spectra with a decreasing cooling rate. To determine the reason behind this difference between Na and Ca ions, their partial charges were evaluated. This is because Na ions are expected to be associated with BOs rather than NBOs as charge compensators by approaching the rings in the further relaxed SLS glass, as discussed in M. Shimizu, et al [11]. The partial charge was obtained by Lowdin charge density analysis using the projwfc.x calculation in QE. The average partial charges of the Na ions were 0.79745e, 0.79432e, and 0.79494e in the models obtained at cooling rates of 0.1, 0.01, and 0.001 K/ps, respectively, and those of Ca ions were 1.7615e, 1.7653e, and 1.7652e, respectively. Accordingly, Na decreased and Ca increased the partial charges as the glass structures were further relaxed. This is consistent with the fact that Na ions are surrounded by more BOs, whereas Ca ions interact with more NBOs in SLS glass, which has a lower fictive temperature [12]. This is because the Raman scattering intensity is higher for more negatively charged ions, as shown in Fig. S6. Na decreased the intensity in its decomposed Raman spectra, whereas Ca increased its intensity as the relaxation proceeded further, as illustrated in Fig. 10. In other words, the variation in the Si-O-Si angles influenced the changes in the decomposed Raman spectra of the modifiers by affecting the ion locations.

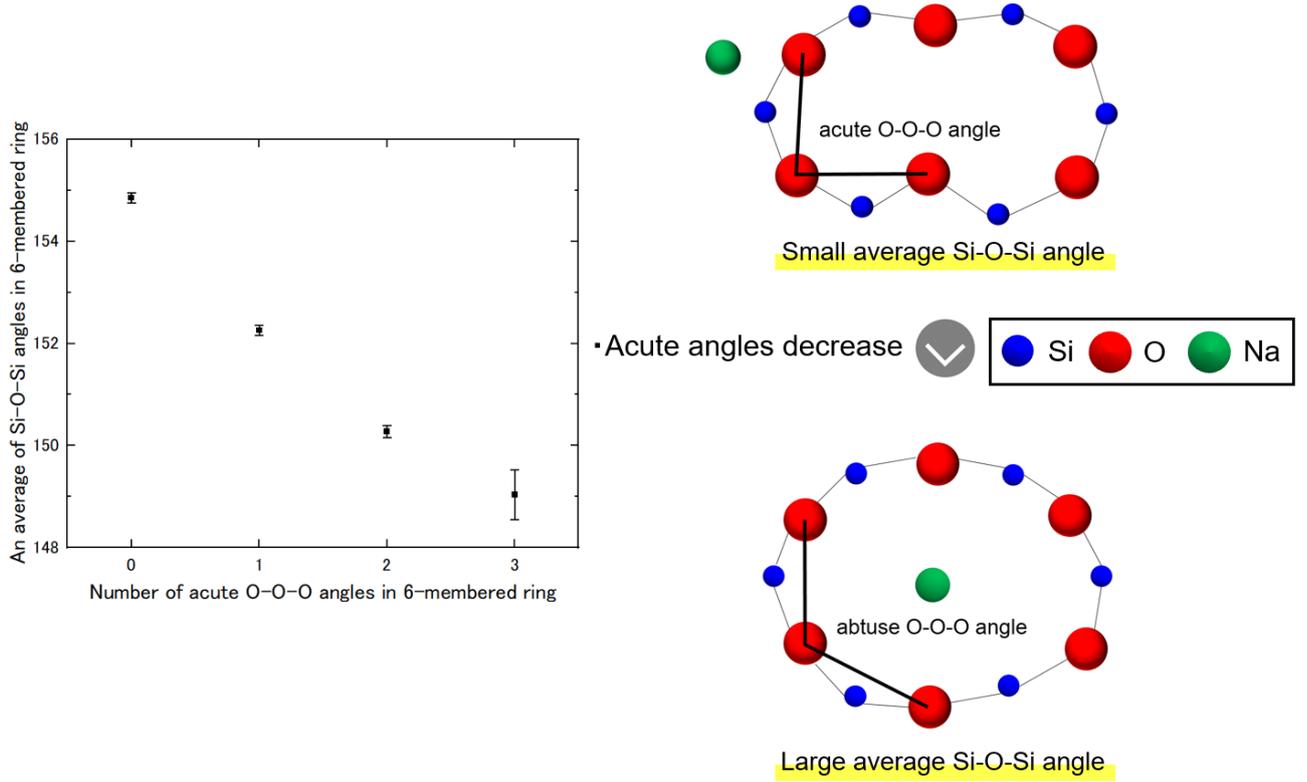

FIG. 14. Correlation between acute O-O-O angles and average Si-O-Si angles in six-membered rings. The data presents the summary of the glasses obtained by cooling rates of 10, 1, 0.1 and 0.01 K/ps. For three acute angles, the error bar is large due to the small number of such rings.

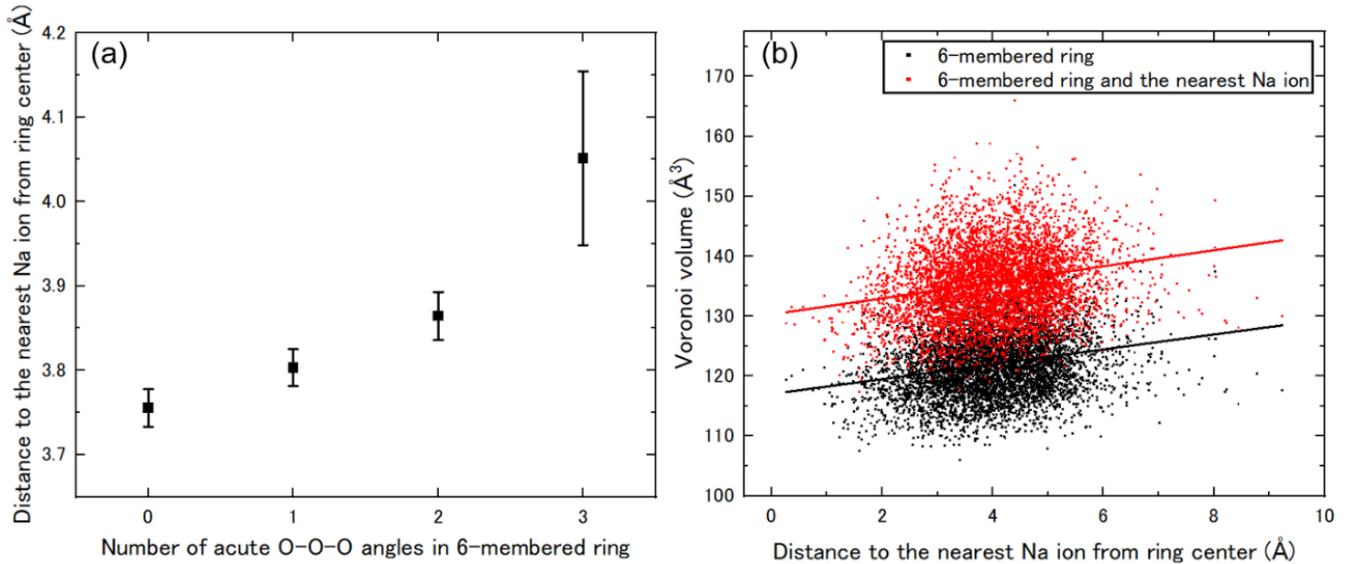

FIG. 15. (a) The relationship between the number of acute O-O-O angles in a six-membered ring and the distance of the nearest Na ion from the ring center. (b) Plots of the distance of the nearest Na ion from the ring center and the Voronoi volume of a six-membered ring and the sum of the six-membered ring and the Na ion.

## V.  CONCLUSION

In this study, experimental and computational investigations were conducted to determine the origins of structural relaxation in SLS glass below the glass transition temperature. By comparing the experimental and theoretical Raman spectra, the structural relaxation below the glass transition temperature is characterized by the reaction $Q_2+Q_4 \rightarrow 2Q_3$ in the early stage. Conversely, the variation in the Si-O-Si angle was attributed to slow relaxation in the later stage.

Based on the detailed analyses done by decomposing the theoretical Raman spectra and considering the ion types and their environments, the following were identified:

1. Oxygen dominates the Raman spectra at frequencies higher than 900 $cm^{-1}$.
2. The Raman band at around 1050 $cm^{-1}$ corresponds to the stretching and bending vibrational modes of BO consisting of small Si-O-Si angles.
3. As the fictive temperature decreases, the 1050 $cm^{-1}$ band area decreases because further structural relaxation widens the small Si-O-Si angles, which reduces the intensity of the 800 $cm^{-1}$ band due to the slower quenching.
4. The six-membered rings, whose acute angles were ameliorated by slower quenching, attracted Na ions, resulting in slow volume relaxation at the later stage.
5. In addition, positional changes in the Na ions decreased their partial charges, resulting in an increase in the intensity of the decomposed Raman spectra of Na as the cooling rate decreased.

Our extensive theoretical considerations unraveled the detailed mechanisms of short- and long-term relaxations in SLS glass and, thus, provided crucial knowledge for controlling the structural shrinkage, which often becomes problematic in industrial applications of oxide glasses.

**SUPPLEMENTARY MATERIAL**
See the supplementary materials for Raman spectra after annealing $T_g$+30 K, $Q_n$-O-$Q_m$ analysis, decomposed Raman spectra, Lowdin charge, the relation between Si-O-Si and O-O-O angle.

**ACKNOWLEDGEMENTS**
Computation time was provided by the Super Computer System at the Institute for Chemical Research, Kyoto University.

**AUTHOR DECLARATIONS**
Conflict of Interest
The authors declare that there is no conflict of interest.

**Author Contributions**
T.S., Y. H., M. S., K. M. formulated this project; T. S. performed the experiment, simulation, analysis, and consideration; S. U. reviewed and edited the paper; T. S. and M. S. wrote the paper with the cooperation of all authors. Correspondence and requests for material should be addressed to M. S.

**DATA AVAILABILITY**

The data that supports the findings of this study are available within the supplementary material.

**Supplementary materials**

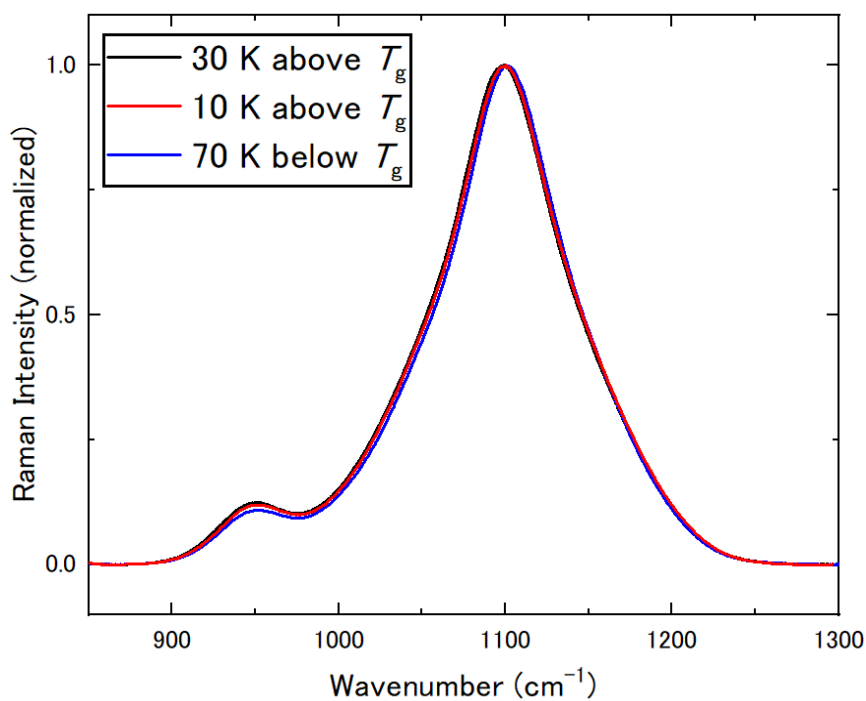

FIG. S1. Raman spectra annealing at 30 K above $T_g$ ($T_g/T$ = 0.964) for 2 h, 10 K above $T_g$ for 2 h, and 70 K below $T_g$ for 1200 h.

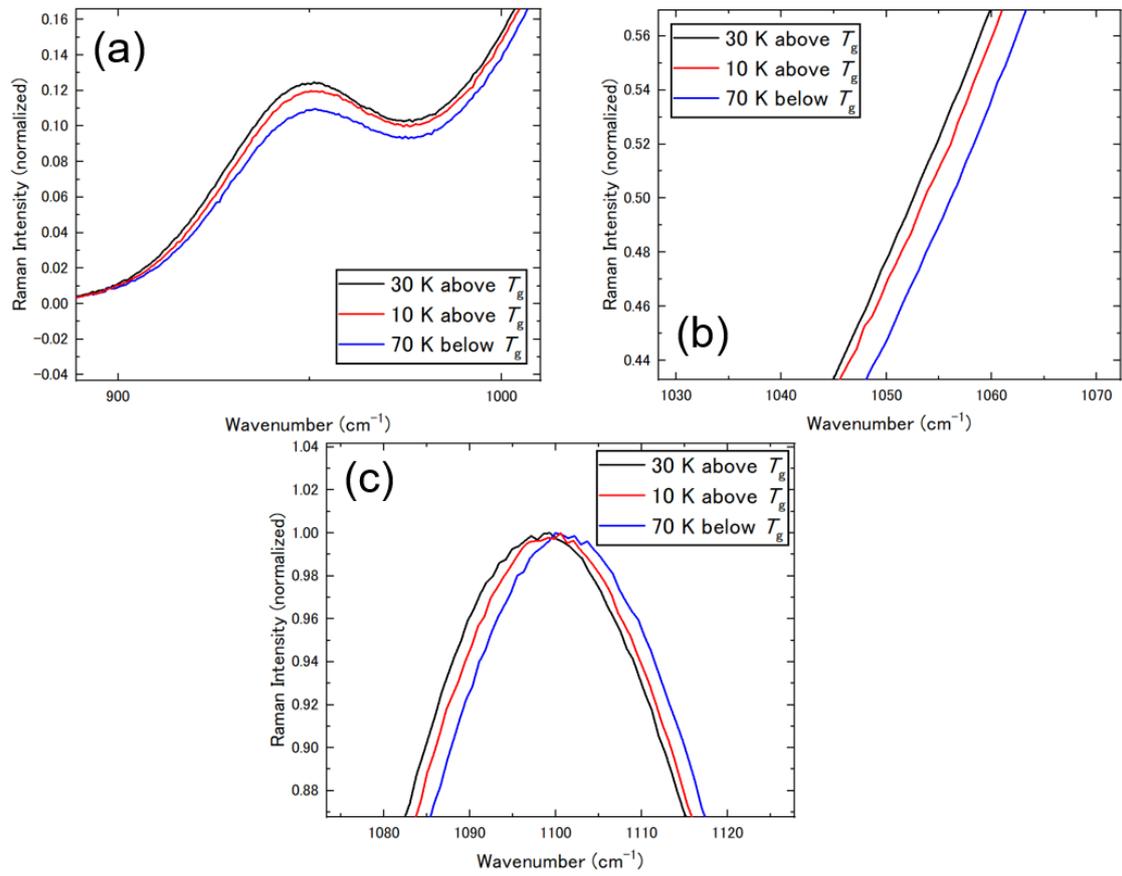

FIG. S2. Extended figures of FIG. 6 at (a) 950, (b) 1050, (c) 1100 cm$^{-1}$.

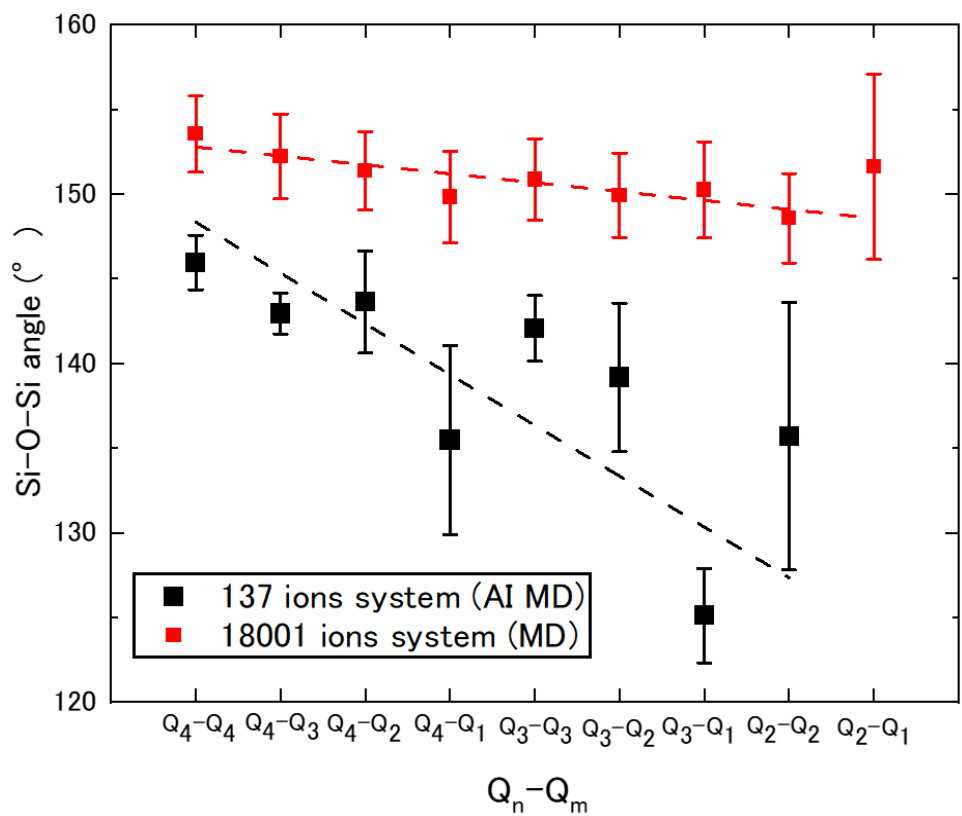

FIG. S3. The relationship between the $Q_n$-O-$Q_m$ species and the size of the Si-O-Si angle in both systems.

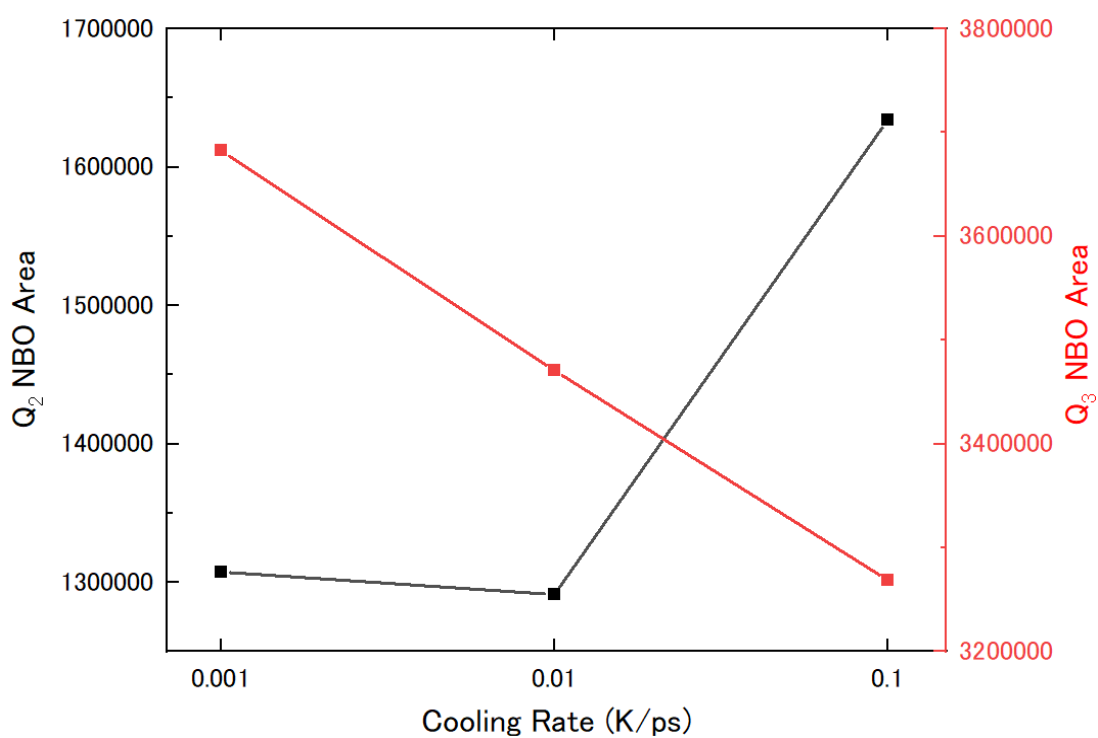

FIG. S4. The cooling rate dependence of the peak area of $Q_2$-NBO (black) and $Q_3$-NBO (red).

**Decomposition of calculated Raman spectra into each vibrational mode**

The decomposition of the Raman spectra of BOs and NBOs by the vibrational mode is shown in Figs. S5(a) and (b). From Fig. S5 (a), we analyzed the BO motions as rocking, bending, and stretching vibrations. For the BO and NBO, the relationship between the phonon eigenvectors and atomic coordinates was used to decompose each vibrational mode. Rocking, bending, and stretching vibrations, respectively, exhibited intensity it the high-frequency region. This is consistent with the VDOS results of previous studies [26]. Comparing each vibration mode, the bending mode is the only one with an intensity at 750 cm$^{-1}$. The peak at 750 cm$^{-1}$ (800 cm$^{-1}$ in the experiment), which was discussed earlier, was attributed to the bending vibration. Furthermore, the vibration mode in the high-frequency region is a combination of stretching and bending vibrations. From Fig. S5 (b), the vibration mode of the NBO in the high-frequency region is attributed to the stretching vibration. Thus, the Raman spectra in the high-frequency region are predominantly influenced by the stretching mode of NBO, and the stretching and bending modes of BO overlap.

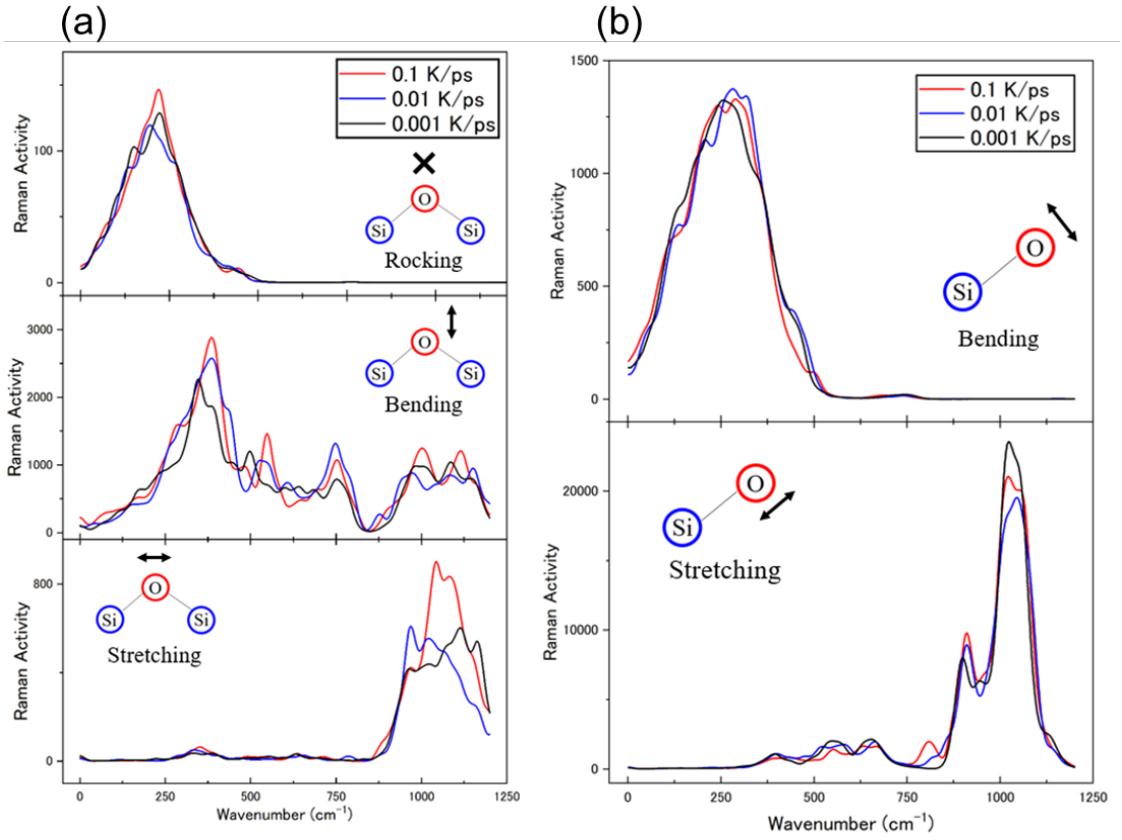

FIG. S5. The decomposition of (a)BOs' and (b) NBOs' Raman spectra by vibrational mode.

**Lowdin charge analysis**

This section describes the relationship between the electrical properties of the chemical species and the Raman spectra. A Lowdin charge density analysis was performed to understand the electrical properties of the SLS glasses. The average partial charge of each ionic species was calculated from the Lowdin charge and number of valence electrons (Table S1). To the best of our knowledge, this is the first time that low-charge-density analysis has been applied to silicate glasses.

We examined the partial charge of Si for each $Q_n$ species. The average charges of Si-$Q_n$ (i.e., Si bonded to n BO) in the SLS glasses were 1.787, 1.698, 1.616, and 1.551 $e$ respectively. It is understood that Si becomes more negatively charged as it bonds with NBO.

For O, the average charges of the NBOs and BOs were -0.9926 and -0.8198 $e$, respectively. Because the NBO has a more negative partial charge, electrostatic repulsion occurs between the electrons. This repulsion causes the electron cloud to expand, which increases the polarizability change due to the electric field of light. Therefore, we conclude that the Raman scattering intensity of NBOs is higher than that of BOs in the high-frequency band (see Fig. 11). The partial charges were calculated by focusing on the BO (i) Si-O-Si angle and (ii) $Q_n$ species on both sides. The correlation between the Si-O-Si angle and Lowdin charge shows that BOs with smaller Si-O-Si angles have more negative partial charges. We conclude that this is because BOs with small Si-O-Si angles tend to bond with negatively charged Si, such as $Q_2$-Si and $Q_1$-Si, on both sides. Figure 12 was normalized per unit, and the relationship between the maximum intensity and the average partial charge of BO was investigated using Fig. S6. We observed a strong negative correlation between the partial charge and the Raman spectral peak intensity. In other words, BOs with small Si-O-Si angles are more negatively charged and thus have a larger Raman scattering cross-section.

In summary, the results of the Lowdin charge density analysis revealed differences in the electrical properties of the species in the SLS glasses. Furthermore, the more negatively charged chemical species contributed more to the intensities of the calculated Raman spectra. From the above, it can be concluded that the peak at 1050 cm$^{-1}$ comes from BOs with a small Si-O-Si angle and more negative partial charges, resulting in larger Raman scattering.

|  | Partial charge (*e*) |  | Partial charge (*e*) |
| --- | --- | --- | --- |
|  |  | BO |  |
| Si-$Q_4$ | 1.787 (0.0161) | 110-120 | -0.8282 (0.0123) |
| Si-$Q_3$ | 1.698 (0.0170) | 120-130 | -0.822 (0.0173) |
| Si-$Q_2$ | 1.616 (0.0177) | 130-140 | -0.8207 (0.0142) |
| Si-$Q_1$ | 1.551 (0.0254) | 140-150 | -0.8195 (0.0136) |
|  |  | 150-160 | -0.8183 (0.0111) |
| NBO | -0.9926 (0.0232) | 160-170 | -0.8178 (0.0094) |
| BO | -0.8198 (0.0136) | 170-180 | -0.8175 (0.0083) |
|  |  |  |  |
| NBO-$Q_3$ | -0.9844 (0.0196) | $Q_4$-$Q_4$ | -0.8111 (0.0090) |
| NBO-$Q_2$ | -1.00464 (0.0211) | $Q_4$-$Q_3$ | -0.8189 (0.0112) |
| NBO-$Q_1$ | -1.026 (0.0220) | $Q_4$-$Q_2$ | -0.8276 (0.0139) |
|  |  | $Q_4$-$Q_1$ | -0.8247 (0.0155) |
|  |  | $Q_3$-$Q_3$ | -0.8278 (0.0133) |
|  |  | $Q_3$-$Q_2$ | -0.8374 (0.0152) |
|  |  | $Q_3$-$Q_1$ | -0.8407 (0.0209) |
|  |  | $Q_2$-$Q_2$ | -0.8473 (0.0100) |

TABLE S1. The average partial charges of each ionic species (O and Si) calculated by Lowdin charge density analysis. The values given in parentheses are the standard deviations of their distribution.

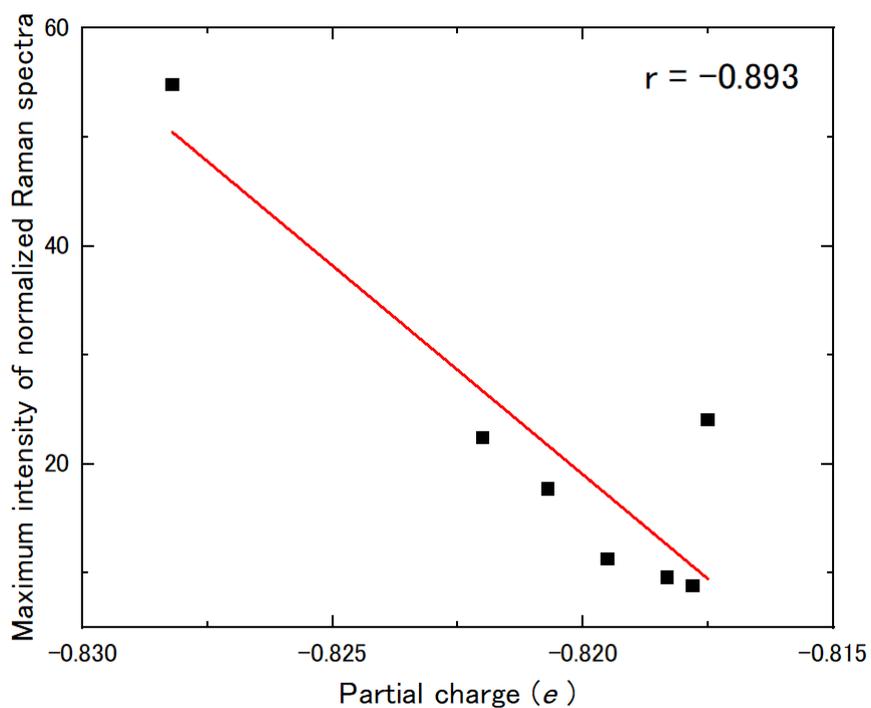

FIG. S6. Correlation between an average BO partial charge and maximum intensity of decomposed Raman spectra per unit of Si-O-Si angles. Calculated Raman spectra were averaged for all cooling rates.

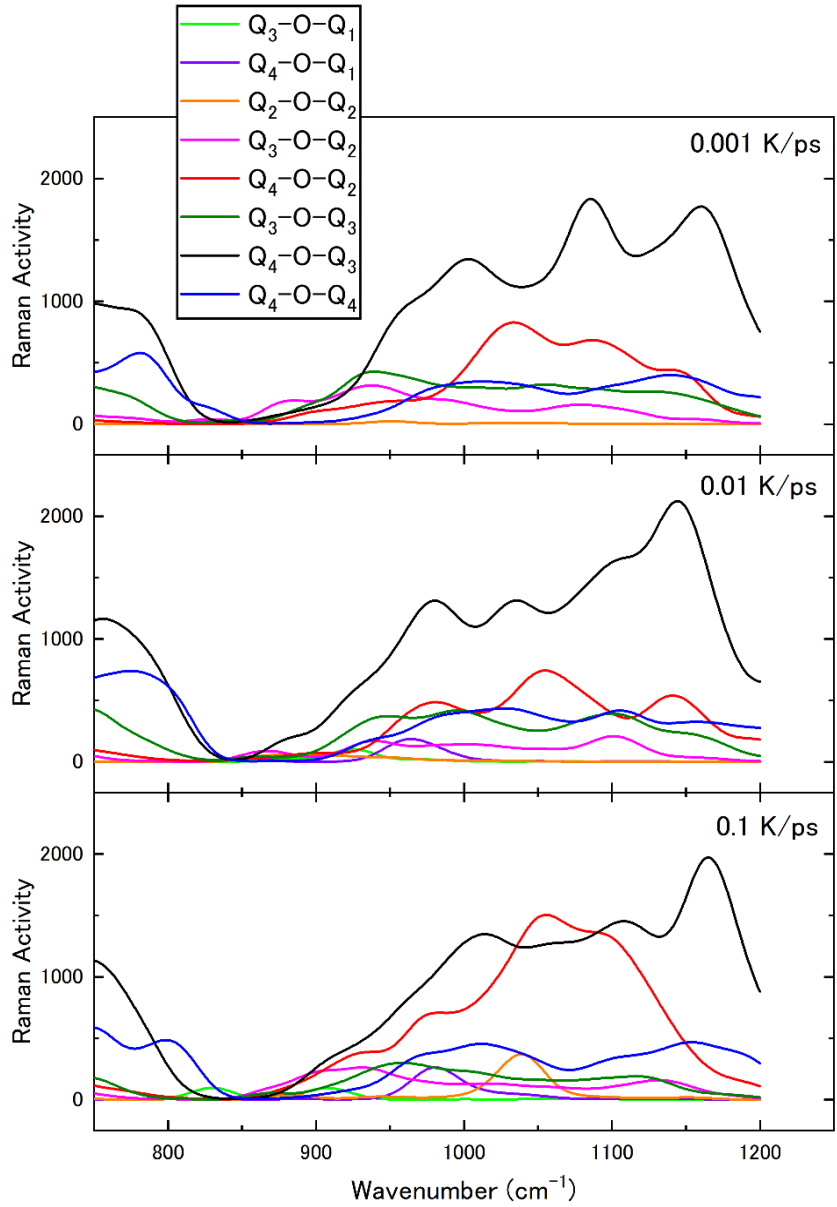

FIG. S7. Decomposed Raman spectra of BOs focusing on $Q_n$-O-$Q_m$ for each cooling rate.

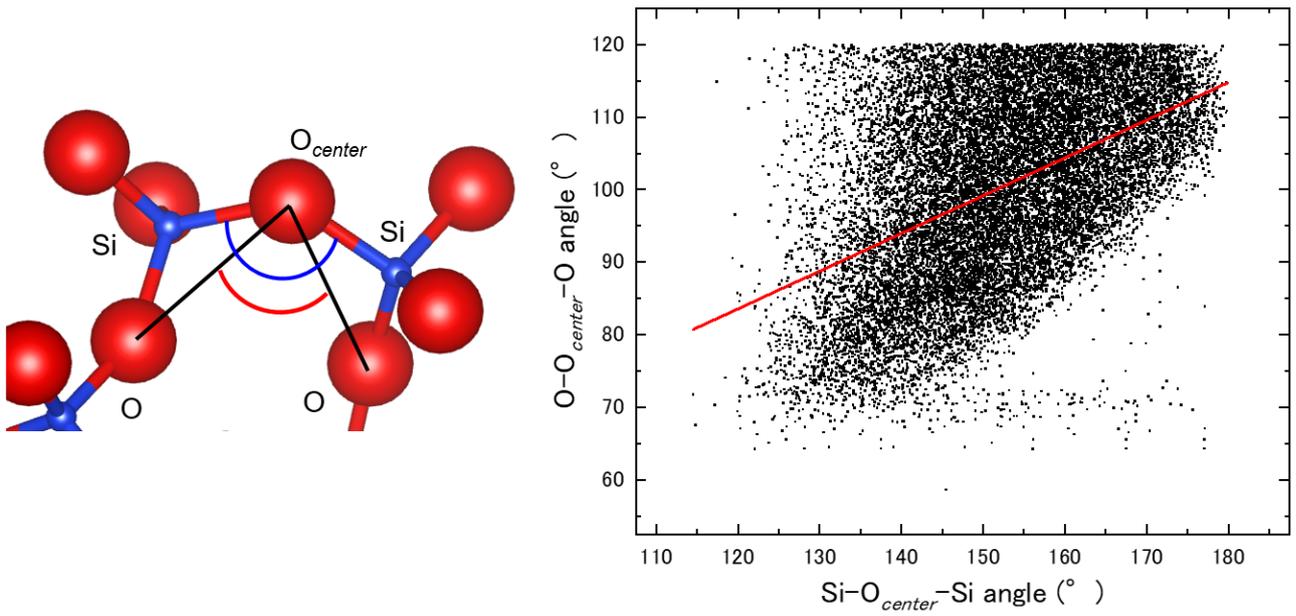

FIG. S8. Relationship between Si-$O_{center}$-Si angle and O-$O_{center}$-O angle in 6-membered ring. The data presents the summary of the glasses obtained by cooling rates of 10, 1, 0.1 and 0.01 K/ps in 18001 ions system.